\documentclass[aps,pra,twocolumn,10pt,superscriptaddress]{revtex4-2}
\pdfoutput = 1
\usepackage{svg}
\usepackage{times,bbm,amsmath,amssymb,amsthm}
\usepackage{epsfig,color}
\usepackage{xcolor}
\usepackage[colorlinks=true,citecolor=blue,linkcolor=blue]{hyperref}
\hypersetup{
   allcolors  = {blue},
}
\usepackage{cleveref}
\usepackage{subfiles}
\usepackage{float}
\usepackage{verbatim}
\usepackage[greek,english]{babel}
\usepackage{thumbpdf,enumerate}
\usepackage{booktabs}
\usepackage{sidecap}
\usepackage[scaled=.8]{couriers}    
\usepackage{pstricks}
\usepackage{multirow}
\usepackage{placeins}
\usepackage{relsize}  
\usepackage{pst-grad,bm}
\usepackage{epigraph}
\usepackage{longtable}
\usepackage{booktabs}
\usepackage{bm}

\usepackage{soul}
\usepackage{ulem} 
\usepackage{acronym}
\usepackage{physics}
\usepackage{tikz}
\newsavebox{\imagebox}

\begin{document}

\title{Quantum Analog-to-Digital Converter for Phase Estimation}

\author{Eugenio Caruccio}
\affiliation{Dipartimento di Fisica, Sapienza Universit\`{a} di Roma, Piazzale Aldo Moro 5, I-00185 Roma, Italy}

\author{Simone Roncallo}
\affiliation{Dipartimento di Fisica, Universit\`{a} di Pavia, via A. Bassi 6, I-27100 Pavia, Italy}

\author{Valeria Cimini}
\affiliation{Dipartimento di Fisica, Sapienza Universit\`{a} di Roma, Piazzale Aldo Moro 5, I-00185 Roma, Italy}

\author{Riccardo Albiero}
\affiliation{Istituto di Fotonica e Nanotecnologie, Consiglio Nazionale delle Ricerche (IFN-CNR), Piazza Leonardo da Vinci, 32, I-20133 Milano, Italy}

\author{Ciro Pentangelo}
\affiliation{Ephos, Viale Decumano 34, I-20157 Milano, Italy}

\author{Francesco Ceccarelli}
\affiliation{Istituto di Fotonica e Nanotecnologie, Consiglio Nazionale delle Ricerche (IFN-CNR), Piazza Leonardo da Vinci, 32, I-20133 Milano, Italy}

\author{Giacomo Corrielli}
\affiliation{Istituto di Fotonica e Nanotecnologie, Consiglio Nazionale delle Ricerche (IFN-CNR), Piazza Leonardo da Vinci, 32, I-20133 Milano, Italy}

\author{Roberto Osellame}
\affiliation{Istituto di Fotonica e Nanotecnologie, Consiglio Nazionale delle Ricerche (IFN-CNR), 
Piazza Leonardo da Vinci, 32, I-20133 Milano, Italy}

\author{Nicol\`{o} Spagnolo}
\affiliation{Dipartimento di Fisica, Sapienza Universit\`{a} di Roma, Piazzale Aldo Moro 5, I-00185 Roma, Italy}

\author{Lorenzo Maccone}
\affiliation{Dipartimento di Fisica, Universit\`{a} di Pavia, via A. Bassi 6, I-27100 Pavia, Italy}

\author{Chiara Macchiavello}
\email{chiara.macchiavello@unipv.it}
\affiliation{Dipartimento di Fisica, Universit\`{a} di Pavia, via A. Bassi 6, I-27100 Pavia, Italy}

\author{Fabio Sciarrino}
\email{fabio.sciarrino@uniroma1.it}
\affiliation{Dipartimento di Fisica, Sapienza Universit\`{a} di Roma, Piazzale Aldo Moro 5, I-00185 Roma, Italy}


\begin{abstract}
Traditional quantum metrology assesses precision using the figures of merit of continuous-valued parameter estimation. Recently, quantum digital estimation was introduced: it evaluates the performance information-theoretically by quantifying the number of significant bits of the parameter, redefining key benchmarks like the Heisenberg bound. Here, we report the first experimental realization of a Quantum Analog-to-Digital Converter for quantum metrology, that takes an input continuous parameter and outputs a bit string, using an advanced photonic platform, comprising a fully reconfigurable integrated circuit and a quantum dot source of highly indistinguishable photons. We implement a protocol for digital phase estimation that is capable of surpassing the standard quantum limit, through the simultaneous use of different entangled state resources. We tackle experimental imperfections by employing machine learning techniques for noise deconvolution and estimation process refinement. Our protocol is experimentally benchmarked against classical strategies via the number of recoverable bits of the unknown parameter. Our results open new perspectives for future implementation of quantum digital estimation strategies.

\end{abstract}
\maketitle

Quantum sensors have a fundamental role in several applications, such as atomic clocks and precision timekeeping \cite{robinson2024direct,nichol2022elementary,pedrozo2020entanglement,appel2009mesoscopic}, magnetic field tracking \cite{craigie2021resource,bonato2017adaptive}, and quantum networks \cite{liu2021distributed,liu2024quantum,malia2022distributed, kim2024distributed, malitesta2023distributed}. Up to now, such sensors have been used as analog devices. Indeed, quantum metrology protocols mainly focus on continuous variable measurements, exploiting quantum effects to enhance estimation precision over their classical counterparts \cite{giovannetti2006quantum,giovannetti2011advances}. 
With recent advancements in quantum computing technologies, it becomes interesting to investigate quantum sensors as digital instruments, involving measurements that yield a finite number of bits of information. A key question is whether the advantages, achieved by analog protocols, persist in such a digital framework. Benchmarking requires different figures of merit to evaluate the quality of the parameter estimation in this framework.
As discussed in \cite{hassani2017digital}, the mutual information quantifies the amount of information two random variables have on one another, namely the shared information between the measurement outcomes and the unknown parameter.

In quantum metrology, the standard approach focuses on the scaling of the root mean squared error in the number of resources $N$, and compares it with two fundamental bounds: the standard quantum limit, which quantifies the bound for classical strategies ($1/\sqrt{N}$), and the Heisenberg limit, which identifies the optimal scaling for quantum strategies ($1/N$) \cite{giovannetti2006quantum,giovannetti2011advances,paris2004quantum,cimini2023experimental,higgins2007entanglement,slussarenko2017unconditional,zhou2018heisenberg,valeri2020review}. Digitally, these bounds translate into a mutual information that scales as $\frac{1}{2}\log_2 N$ and as $\log_2 N$ for the classical and quantum scenarios, respectively \cite{hassani2017digital,lu2024number,gorecki2024mutual}.

In this work, we experimentally implement a protocol for the digital quantum estimation of an optical phase, using an integrated photonic circuit \cite{wang2020integrated,pelucchi2022potential}. Our approach leverages the quantum phase estimation algorithm \cite{cleve1998phase} as a Quantum-Analog-to-Digital Converter (QADC) for phase estimation, that encodes the analog phase in three bits of information. As probes, we generate a sequence of distinct resource states: a 4-photon Greenberger–Horne–Zeilinger (GHZ) state, a 2-photon entangled state, and a single-photon state. Conditional operations are subsequently applied to each probe, based on the outcomes of the preceding measurements. To evaluate the protocol performance in terms of mutual information and estimation accuracy, we implement a noise mitigation strategy, taking inspiration from a series of previous works that employed machine learning techniques for single and multiparameter estimation problems \cite{cimini2019calibration,cimini2021calibration,cimini2023deep,cimini2024variational}. The approach used in this work is different. Indeed, we use a specifically tailored denoising autoencoder, trained via the ideal probability distribution at the output of the protocol, to deconvolve and reduce the experimental noise. The autoencoder output is subsequently processed by a feed-forward neural network to mitigate the bias inherent in the estimation process. Within this framework, we demonstrate that the quantum strategy yields a higher number of useful bits compared to the classical interferometric estimation, using an equivalent amount of resources.

\section*{Results}

\subsection*{Protocol}

\begin{figure}[ht!]
\includegraphics[width=0.99\columnwidth]{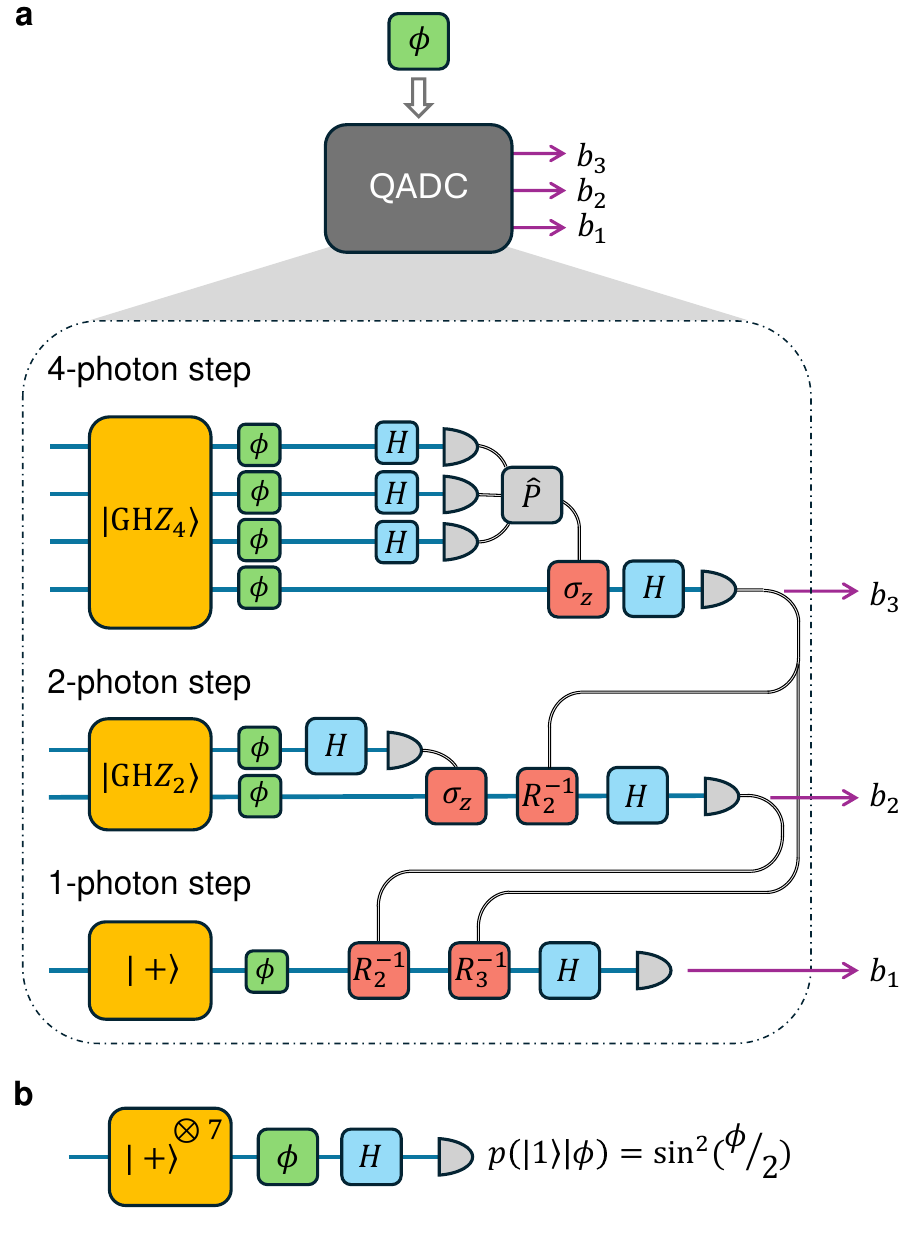}
    \caption{\textbf{QADC phase estimation protocol.} \textbf{a} Quantum digital phase estimation protocol consisting of three stages involving 4, 2 and 1 qubit states, respectively. A unitary operation encodes the phase $\phi$ in each mode, followed by Hadamard gates and a $\sigma_z$ operation, conditioned on the measurement outcomes on the first modes. Depending on the results obtained in the 4-qubit and 2-qubit steps, controlled phase gates $R_{l}^{-1}$ are taken. The bit string $\{b_1,b_2,b_3\}$ is obtained at the output. \textbf{b}, Classical interferometric estimation.}
    \label{fig:conceptual}
\end{figure}

We first briefly review the parallel quantum phase estimation algorithm  \cite{hassani2017digital}. Consider a group of $t$ probes, respectively made of $k = 1,2,\ldots,2^{t-1}$ qubits and prepared in the state $\ket{\text{GHZ}_{k}} = (\ket{0}^{\otimes k} + \ket{1}^{\otimes k})/\sqrt{2}$. Encode the optical phase $\phi \in [0,2\pi)$ in $U(\phi) = \ket{0}\!\bra{0} + e^{\imath \phi}\ket{1}\!\bra{1}$, evolving each probe with $U^{\otimes k}$. Apply a chain of $k-1$ subsequent CNOTs, after which, the $k$th qubit evolves into $(\ket{0}+e^{\imath k\phi}\ket{1})/\sqrt{2}$. Let $\ket{m}$ label the computational basis of the register made by the least significant qubit of each probe, with size $t$ and state $\sum_{m} e^{\imath m \phi}\ket{m}/\sqrt{2^t}$. Taking the inverse quantum Fourier transform (QFT$^\dagger$) yields $\ket{b_1 b_2 b_3}$. Measuring in the computational basis provides the binary expansion of the optical phase as $\phi = 2\pi(b_1/2 + b_2/4 + b_3/8)$, and thus the protocol acts as a QADC for the estimation process. We discuss a specific implementation of such protocol, tailored to the photonic platform design. In this case, the CNOTs chain is achieved by a sequence of Hadamard gates, followed by a controlled $\sigma_z$ operation. The QFT$^\dagger$, instead, is implemented without SWAPs \cite{nielsen2010quantum}, dividing the experiment in three ordered parts: $4$-photon, $2$-photon and $1$-photon steps, whose outputs control the rotations $R_l^{-1} = U(-2\pi/2^{l})$ (see Fig. \ref{fig:conceptual}). 

\textbf{$\bm{4}$-photon} Consider the $4$-photon state $\ket{\text{GHZ}_{4}}$, evolved by the phase transformation into $(\ket{0000} + e^{\imath 4\phi}\ket{1111})/\sqrt{2}$. Apply Hadamard gates to the first three photons, then measure their parity $\hat{P}$ in the computational basis, which reads $1$ or $0$ for bit strings that contain an odd or even numbers of $1$s, respectively. When the parity gives $0$, the fourth photon is in $(\ket{0}+e^{\imath 4\phi}\ket{1})/\sqrt{2}$, otherwise apply $\sigma_z$, obtaining the same result. This procedure, which can be implemented in post-selection, is equivalent to the CNOTs chain. Finally, apply a Hadamard gate and a measurement in the computational basis. The output returns $b_3$, the least significant bit of the binary expansion of $\phi$.

\textbf{$\bm{2}$-photon} Consider the $2$-photon state $\ket{\text{GHZ}_{2}}$. Due the phase shift, this evolves into $(\ket{00} + e^{\imath 2\phi}\ket{11})/\sqrt{2}$. Apply a Hadamard gate to the first photon, followed by a computational basis measurement. When this returns $0$, the other photon is projected into $(\ket{0}+e^{\imath 2\phi}\ket{1})/\sqrt{2}$, otherwise apply $\sigma_z$. We implement the QFT$^\dagger$ adding a rotation $R_2^{-1}$, controlled by the value of $b_3$ and followed by a Hadamard gate. A measurement in the computational basis yields $b_2$.

\textbf{$\bm{1}$-photon} Consider the $1$-photon state $\ket{+}$. The phase shift gives $(\ket{0} + e^{\imath \phi}\ket{1})/\sqrt{2}$. Apply two subsequent rotations, $R_2^{-1}$ and $R_3^{-1}$, controlled by $b_2$ and $b_3$, respectively. Finally, apply a Hadamard gate and a measurement in the computational basis. This gives $b_1$, the most significant bit of the binary expansion of $\phi$.

The classical counterpart of our protocol consists in an entanglement-free interferometric estimation. Consider $2^{t}-1$ qubits, all prepared in $\ket{+}$. After the phase shift, each qubit evolves into $(\ket{0} + e^{\imath \phi}\ket{1})/\sqrt{2}$. Apply a Hadamard gate, followed by a computational basis measurement. For each qubit, $p(\ket{1}\!|\phi) = \sin(\phi^2/2)$. The optical phase reads $\phi = 2 \arcsin(N_1/(2^{t}-1))$, where $N_1$  is the number of $1$s in the output bit string.

\subsection*{Experimental implementation}

\begin{figure*}[ht!]
\includegraphics[width=0.99\textwidth]{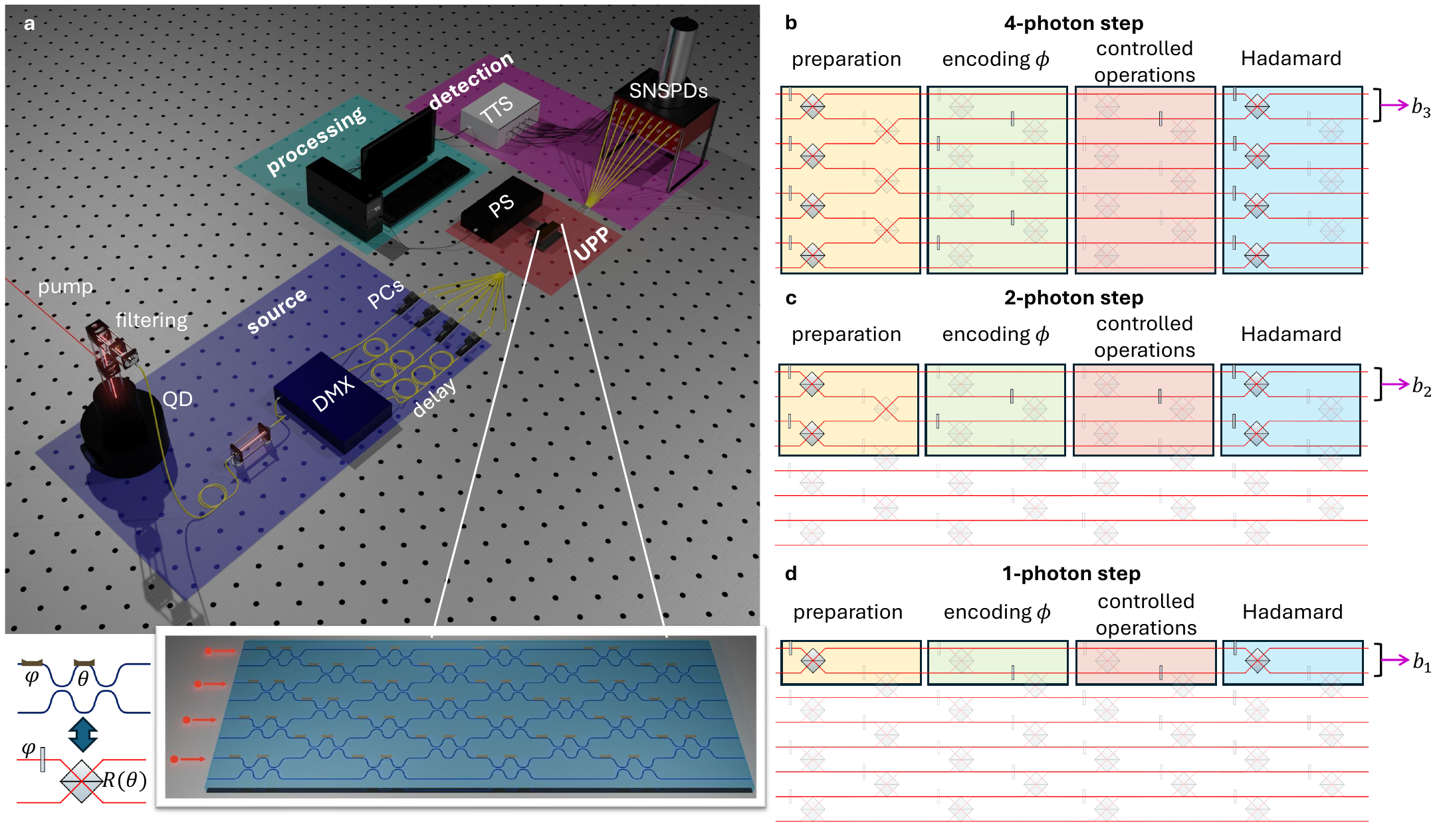}
\caption{\textbf{Experimental setup for the QADC.} \textbf{a}, Schematic view of the experimental setup comprising the quantum-dot source, a time-to-spatial demultiplexing module, coupling to the integrated circuit, and the detection stage. Circuit programming and data acquisition are orchestrated by a processing unit. Inset: internal structure of the integrated device, comprising 28 fully reconfigurable Mach-Zehnder interferometer, acting as elementary cell, which corresponds to a fully-programmable beam-splitter transformation on the modes. Legend: QD - quantum-dot, PCs - polarization controllers, PS - power supply, UPP - universal photonic processor, SNSPDs - superconducting nanowire single-photon detectors, TTS - time-tagging system. \textbf{b}-\textbf{d}, Circuit programming for the different steps of the protocol. Orange: layers used for the state preparation. Green: layers used for phase encoding. Red: layers used for the $\sigma_{z}$ and $R_{l}^{-1}$ operations. Cyan: layers used for the Hadamard operations.}
\label{fig:setup}
\end{figure*}

The experimental apparatus, for the implementation of the digital estimation algorithm, is based on a photonic hybrid platform, depicted in Fig. \ref{fig:setup}a. The exploited photon source is a quantum-dot (QD) semiconductor, emitting a train of single photons. A 79 MHz pulsed laser is injected in an optical setup tailored to double the pulse repetition rate up to 158 MHz, and then is subsequently employed to excite (pump) the QD source at the same wavelength of the energy transition (RF regime) \cite{somaschi2016near, gazzano2013bright}. Single photons are properly discriminated by the pump laser, using a cross-polarization filtering scheme. A time-to-spatial demultiplexer (DMX), based on an acousto-optic effect \cite{pont2022quantifying, rodari2024semi}, spatially splits the train of photons into $n$ channels, every 180 ns. Finally, the photon streams are temporally synchronized by fiber delays, and a set of polarization controllers (PCs) compensate for the relative polarization among the channels. In this way, a $n$-photon Fock state is produced in the photon spatial degree of freedom. Specifically, the photon numbers required for this experiment are $n=1,2,4$. Subsequently, the multi-photon state is injected into a universal photonic processor (UPP). The UPP consists of a fully programmable $8$-mode photonic integrated circuit \cite{pentangelo2022universal, pentangelo2024high} and a power supply (PS), which provides electrical currents to the chip and allows implementing arbitrary unitary transformations by actuating multiple thermal phase shifters on top of the photonic circuit. The optical circuit comprises $28$ Mach-Zehnder unit cells, organized according to a universal rectangular layout \cite{clements2016optimal}. The unit cell is a reconfigurable beam-splitter composed of two balanced directional couplers and two reconfigurable phases $\theta\in[0,\pi)$ and $\varphi\in[0,2\pi)$. More specifically, the reflectance $R$ of the reconfigurable beam-splitter is set by tuning $\theta$, while $\varphi$ determines the complex phase of the two-mode transformation. Finally, single photons exiting the interferometer are routed towards a detection stage, which is constituted by superconductive nanowire single-photon detectors (SNSPDs) and a time-tagging system (TTS), having $\sim 42$ ps time resolution, which is capable of registering $n$-fold coincidences detection events. A processing unit is involved in controlling the power supply that sets the currents on the thermo-optic phase shifters, and in recording the detection events coming from the TTS.

The protocol is implemented in the dual-rail encoding \cite{kok2007linear}, where each rail consists of a spatial mode of the device. In such a way, the first qubit is encoded in the logical state $\vert 0 \rangle$ or $\vert 1 \rangle$ depending on the presence of a photon in one of the modes (1,2) of the 8-mode interferometer, the second qubit by a photon in the modes (3,4), and henceforth. The experiment relies on the reconfiguration capabilities of our device, which can be programmed to implement each step of Fig. \ref{fig:conceptual}, where each section of the processor is used to obtain the required functionalities (see Fig. \ref{fig:setup}b-d). (i) As a first step, the input resource states are generated via the first two layers of the interferometer. For instance, $n$-photon GHZ states \cite{pont2022high} are prepared with a first layer of reconfigurable beam-splitters with $R=0.5$, and a second layer where the beam-splitters are programmed in cross-configuration ($R=0$) (orange section in Fig. \ref{fig:setup}b-c). The preparation is performed in a post-selected configuration conditioned to the presence of one photon for each pair of dual-rail encoded modes. Such scheme leads to the generation of states $\vert \tilde{\mathrm{GHZ}}_4 \rangle = (\vert 0101 \rangle + \vert 1010)/\sqrt{2}$ and $\vert \tilde{\mathrm{GHZ}}_2 \rangle = (\vert 01 \rangle + \vert 10)/\sqrt{2}$ for the $4$- and $2$-photon case respectively. Those states are equivalent up to local unitary operations to those discussed in the protocol section. The $1$-qubit state $(\ket{0}+\ket{1})/\sqrt{2}$ (Fig. \ref{fig:setup}d) is generated deterministically with a single reconfigurable balanced beam-splitter. (ii) The phase ($\phi$) is encoded in the second section of the circuit (green in Fig. \ref{fig:setup}b-d), while (iii) the third section implements the controlled phase gates $R_l^{-1}$ and $\sigma_z$ (red in Fig. \ref{fig:setup}b-d). Here, $\sigma_z$ is controlled by the parity of the output bit strings, while $R_{l}^{-1}$ require feed-forward propagation after each step. Such operations are implemented in post-selection by collecting data for all possible configurations, and then by applying post-processing to properly combine such data. (iv) The Hadamard operations $H$ are applied in the final section (cyan in Fig. \ref{fig:setup}b-d). More details on the data analysis can be found in the Supplementary Information.

\begin{figure*}[!htb]
\includegraphics[width=0.99\textwidth]{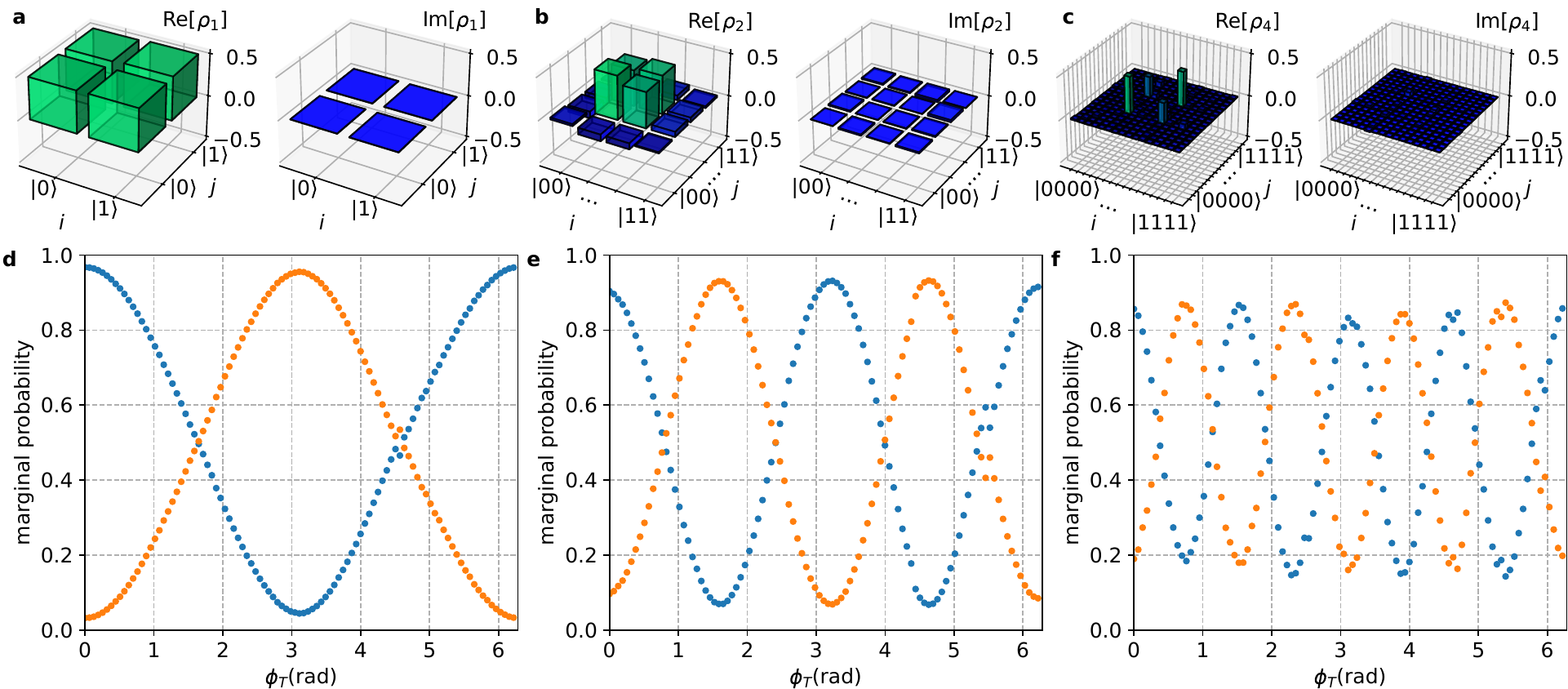}
    \caption{\textbf{Calibration of the measurement outcomes.} \textbf{a}-\textbf{c}, Plots of the experimental tomographies for the 1-, 2-, 4-qubit states employed for the quantum digital protocol.  
    \textbf{d}-\textbf{f}, Plots of the experimentally reconstructed marginal probabilities on the informative qubits, for the 1-, 2- and 4-photon post-selected measurement outcomes of the digital estimation scheme.}
    \label{fig:conceptual2}
\end{figure*}

\subsection*{Experimental results}

\begin{figure*}[!htb]
\includegraphics[width=0.99\textwidth]{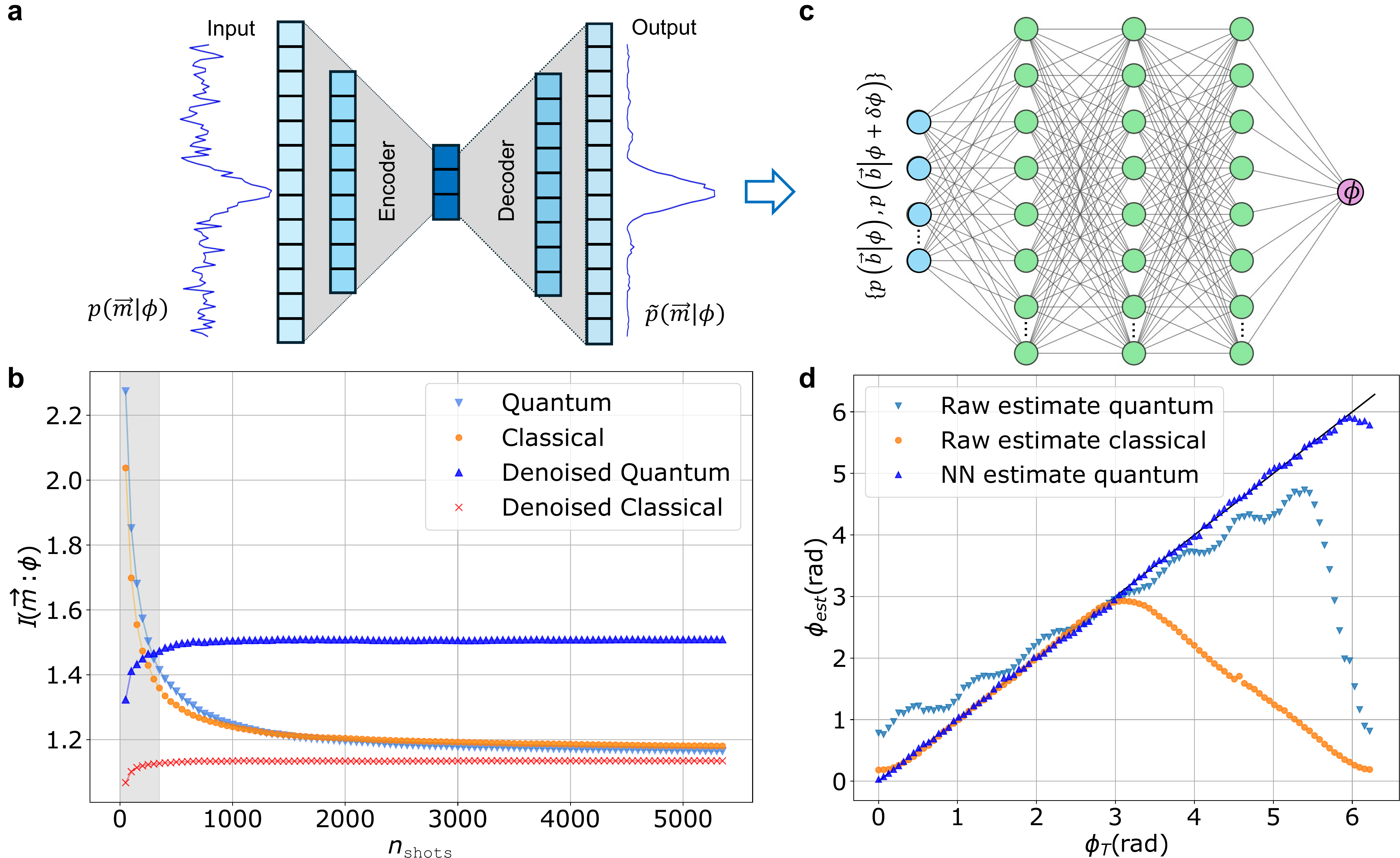}
\caption{\textbf{Denoising autoencoder and feed-forward neural network architectures together with experimental results.} \textbf{a}, Architecture of the denoising autoencoder (DAE) implemented to mitigate the influence of noise on the conditional probabilities $\tilde{p}(\vec{m}|\phi)$, reconstructed from experimental data. \textbf{b}, Comparison of the mutual information computed from experimental data using quantum probes (light blue) against the one retrieved with single-photon probes (orange). The same quantity is then computed using the denoised experimental probabilities of the quantum (blue triangles) and the classical (red crosses) probes. Note that the mutual information becomes a meaningful figure of merit only after accumulating a sufficient number of probes; for this reason, we have shaded (grey) the initial region. \textbf{c}, Feed-forward neural network (NN) architecture used for phase estimation, trained to map the reconstructed conditional probabilities to the true phase values, refining the estimation process.  \textbf{d}, Optical phase estimate $\phi_{\text{est}}$ obtained by applying the quantum digital protocol (light blue) and the classical interferometric strategy (orange) as a function of the true value of the parameter $\phi_T$. The performance of the quantum protocol is further improved by employing the NN able to remove ambiguities.}
\label{fig:results}
\end{figure*}

A thorough characterization of the system operation is conducted by examining the generation of the various resource states. This analysis serves as a crucial preliminary step before implementing the estimation protocol, offering an overview of the generated quantum states and their corresponding measurement outcomes. The results of this characterization are presented in Fig. \ref{fig:conceptual2}. More specifically, in Fig. \ref{fig:conceptual2}a-c we report the tomographic reconstructions of the generated resource states ($\rho_1$, $\rho_2$, $\rho_4$) for the $1$-, $2$- and $4$-photon steps respectively. The fidelities between the reconstructed state and the ideal state are $F^{(1)}=0.9972\pm 0.0001$ for the state $(\ket{0}+\ket{1})/\sqrt{2}$, $F^{(2)}=0.9476\pm 0.0004$ for the state $\vert \tilde{\mathrm{GHZ}}_2 \rangle$, and $F^{(4)}=0.811\pm 0.005$ for the state $\vert \tilde{\mathrm{GHZ}}_4 \rangle$. Such values depend on the different noise sources present in the platform (see Supplementary Information). Additionally, in Fig. \ref{fig:conceptual2}d-f we report the experimentally retrieved marginal probability distributions as a function of the phase $\phi$, where all conditional phases are set to $0$. These probabilities are expected to have a periodic behaviour with phase-period proportional to $1/n$, being $n$ the number of photons.

To validate the protocol experimentally, we collected a dataset comprising a vector $\vec{m}$ of 7 distinct measurement outcomes for $99$ phase values uniformly distributed across the interval $[0,2\pi)$. For each phase value, the measurements are repeated $n^{\mathrm{max}}_{\mathrm{shots}}=5377$ times to ensure the statistical robustness of the obtained results. From the collected data, we reconstruct the conditional probability distribution $p(\vec{m}|\phi)$, required for computing the mutual information. Additionally, we estimate the optical phase using the quantum protocol described, benchmarking its performance against that of the classical strategy shown in Fig. \ref{fig:conceptual}.

For the implementation of the quantum estimation protocol, the phase to be estimated is applied across different modes of the integrated photonic device, depending on the quantum state used, as described above. Specifically, the phase is set on four modes when utilizing the GHZ state, two modes for a two-photon state, and a single mode for single-photon states. Post-selected measurement outcomes are used to retrieve the phase estimates via a digital estimation approach, as previously discussed.
This quantum digital strategy is compared to a classical method, where the device is seeded with single photons, and phase estimation is performed using standard interferometric procedures based on the outcome probabilities of photons in the upper or lower mode. 

Experimental results reveal that for moderate values of $n_{\mathrm{shots}}$ the mutual information obtained with the quantum strategy exceeds that of the classical counterpart, derived from single-photon measurements and the standard interferometric approach. However, as the number of phase repetitions increases, the observed advantage diminishes, highlighting the sensitivity of the quantum probes to experimental imperfections. This effect arises from the noise factors that have a greater impact on the quantum experiment compared to the classical approach. These factors, mentioned previously, include optical losses within the setup and errors in the circuit programming, partial distinguishability of the multi-photon probe states, and fluctuations in the $g(2)$ correlation function during the extended measurement period (see Supplementary Information for numerical simulations discussing impact of the different noise factors).

To mitigate the impact of these experimental imperfections, we exploit a denoising autoencoder (DAE) to process the probabilities reconstructed from raw experimental data. A DAE is a neural network (NN) comprising two components: an encoder, which maps the input data to a compressed latent representation by capturing its essential features, and a decoder, which reconstructs the original data from this representation while discarding irrelevant variations caused by noise. This architecture is particularly well-suited for isolating meaningful features embedded in noisy data \cite{vincent2008extracting}, effectively reducing the impact of experimental imperfections. DAEs have been widely adopted in various unsupervised learning tasks, including image denoising \cite{bajaj2020autoencoders}, anomaly detection \cite{sakurada2014anomaly}, and speech recognition tasks \cite{lu2013speech}, due to their robustness and ability to generalize well to unseen noisy data.

For our application, we train the DAE using ideal conditional probabilities corrupted with Gaussian noise, with the targets set as the clean, noise-free probabilities. This approach forces the encoder to identify invariant features resilient to noise and enables the decoder to reverse the noise effects during reconstruction. We take as a loss function the mean squared error for driving the optimization process, ensuring that the reconstructed probabilities closely match the clean targets. Once trained, the DAE is applied to the experimental probabilities $p(\vec{m}|\phi)$, reconstructed from our raw data, obtaining their denoised version $\tilde{p}(\vec{m}|\phi)$. The scheme representing the action of the DAE is reported in Fig.\ref{fig:results}a.

Phase estimation requires only the probabilities $\tilde{p}(\vec{b}|\phi)$, obtained by marginalizing $\tilde{p}(\vec{m}|\phi)$ to the significant bits $\vec{b} = (b_1, b_2, b_3)$. At the output of the DAE, we compute a vector of $2^3$ elements, representing the frequencies of observing each computational basis state, namely $\tilde{p}(\vec{b}|\phi)$.

To show the effectiveness of the denoising approach, we analyze the mutual information obtained with quantum probes compared to classical ones as a function of the number of detected probes, $n_{\mathrm{shots}}$ reporting the experimental results in Fig.\ref{fig:results}b. After using the denoised probabilities to compute the mutual information, the advantage of the quantum strategy becomes evident regardless of the number of phase repetitions considered.

As a next step, we study the performance of the phase estimation strategy obtained through the described digital protocol. It can be shown that, even under ideal conditions (see Supplementary Information), the retrieved phase estimate exhibits an oscillatory behavior, with values close to the true phase value inspected. For this reason, the digital quantum strategy can be further enhanced, in terms of overall accuracy, by integrating a second feed-forward NN that linearizes the phase estimation process. Specifically, this can be done by training the NN on simulated measurement outcomes corresponding to a range of phase values. These simulations provide a mapping between the measured data and the actual phase values, enabling the network to optimize the estimation process. In this case, the NN is used to learn the complex relationship between the conditional probabilities $p(\vec{b}|\phi)$ and the corresponding phase values, capturing both linear and non-linear dependencies. The training data incorporates noise and variations consistent with experimental conditions, ensuring the network ability to generalize effectively to actual experimental data. Once the NN is trained, it can be directly fed with the experimentally denoised conditional probabilities $\tilde{p}(\vec{b}|\phi)$ to give a refined and more accurate phase estimate as reported in Fig.\ref{fig:results}c. The denoising process effectively restores the quality of the experimental probabilities, allowing for a more accurate computation both of the mutual information and of the phase estimate.

The performance of the estimation strategy obtained with the experimental data are reported in Fig.\ref{fig:results}d. Here, the average phase estimation results over $n_{\mathrm{shots}}$ detected probes for both quantum and classical strategies are shown. A notable distinction between the two approaches arises from the periodicity inherent in the outcome probabilities. Classical probes, constrained by their single-photon nature and standard interferometric techniques, can only resolve phases within the interval $[0,\pi)$. In contrast, entangled quantum probes enable phase discrimination across the entire $2\pi$ range. A significant improvement is obtained when applying the second NN to the denoised probabilities, obtained from the experimental quantum data. Once the NN is trained, it can be directly fed with the experimentally denoised conditional probabilities $\tilde{p}(\vec{b}|\phi)$ to give a refined and more accurate phase estimate. This approach significantly reduces systematic deviations and non-linear distortions inherent in the raw estimation process, providing an unbiased result with no ambiguities or periodicity constraints. The improved estimates obtained using this NN-assisted digital strategy are presented also in Fig.~\ref{fig:results}d, demonstrating indeed enhanced accuracy and consistency compared to the raw estimate. The NN also compensates for imperfections in the experimental setup making the estimation strategy more robust to different sources of noise without requiring the use of additional physical resources, since the NN is previously trained with simulated data.

\section*{Discussion}

In conclusion, we have experimentally implemented a digital quantum protocol for estimating an unknown phase parameter making use of a Quantum Analog-to-Digital Converter. The protocol has been realized by programming a 8-mode fully-reconfigurable universal interferometer to generate and process different entangled resources, with the input ones obtained via a quantum-dot source of highly indistinguishable photons. Using deep learning algorithms (trained on the theoretical probabilities of the ideal protocol), we mitigated noise and refined the estimation analysis, overcoming the mutual information bound of the classical strategies. This means that our protocol can extract more bits of information with respect to a classical approach, by using the same amount of resources (photons and phase shift applications). The potential of QADC is wide and it can be applied to different contexts, for instance, it has been used, only at the theoretical level, as a quantum computational subprimitive in \cite{mitarai2019quantum,schmuser2005quantum}. There, at variance with our scenario, the output of the converter is encoded on qubits. 

Our experiment provides evidence of the capabilities of digital estimation approaches, which have the potential to bridge the gap between quantum sensing and digital signal processing.  Our results pave the way for applications of quantum sensors in all those scenarios where data must be digitalized for real-time processing. 


%

\section*{Methods}

\subsection*{Details on the experimental platform}
In this Section, we provide more details on the experimental platform and on the different parameters characterizing its performance. We first discuss the characterization of the source, then we provide more details on the integrated circuits, and finally, we discuss the observed values of the $n$-fold coincidence rate.

\textbf{Characterization of the single-photon source.} In order to benchmark the performances of the source, it is necessary to characterize different parameters. On a first note, the signal rate obtained at the output of the source is quantified via the brightness parameter, which is mainly driven by the efficiency of QD coupling with the optical cavity and the cross-polarization filtering mechanism through which, in the optimal case, half of the photons are discarded. In our case the brightness parameter is $\sim14\%$. The quality of the generated photons depends on their single-photon purity, and on their mutual indistinguishability. The first aspect corresponds to the need of minimizing any multiphoton component, and can be characterized via the second-order correlation function $g^{(2)}(0)$. This parameter can be measured via a Hanbury-Brown and Twiss interferometer \cite{Brown1954}, and in the present case is found to be $g^{(2)}(0) = (5.321\pm0.001)\times 10^{-3}$, for the $2$-photon experiment, and $g^{(2)}(0) = (5.629\pm0.003)\times 10^{-3}$, for the $4$-photon experiment. Finally, the mutual indistinguishabiity of subsequently emitted photons is a relevant parameter given that the resource states are generated via quantum interference. This parameter can be quantified by reconstructing the Gram-matrix $S_{ij} = \langle \psi_{i} \vert \psi_{j} \rangle$, a $n \times n$ matrix with elements corresponding to the overlap between the states in the photon internal degree of freedom. The moduli can be estimated by measuring the Hong-Ou-Mandel visibilities $V^{\mathrm{HOM}}_{ij}$ of each pair of photons, correcting for multiphoton components \cite{olli21imperfect}, while the non-trivial complex phases are found to be 0 for the employed quantum dot source \cite{rodari2024semi}. In our case, we measured, in the $2$-photon experiment, $V^{\mathrm{HOM}}_{02} = (0.938\pm0.002)$, where the lower indices refer to the input modes. In the $4$-photon experiment, there are $6$ combinations of possible pairs of photon and the HOM visibilities are: $V^{\mathrm{HOM}}_{02} = (0.9334\pm0.0008)$, $V^{\mathrm{HOM}}_{04} =(0.937\pm0.001)$, $V^{\mathrm{HOM}}_{06} =(0.940\pm0.001)$, $V^{\mathrm{HOM}}_{24} =(0.921\pm0.001)$, $V^{\mathrm{HOM}}_{26} =(0.9078\pm0.0009)$, $V^{\mathrm{HOM}}_{46} =(0.918\pm0.002)$.  For a more detailed impact of multiphoton components and partial photon distinguishiability on the protocol performances we refer to the Supplementary Information.

\textbf{Calibration of the reconfigurable universal photonic processor.} The reconfigurable interferometer is a $8 \times 8$ universal processor implemented via the femtosecond laser-writing technique, composed of $28$ Mach-Zehnder interferometers arranged in the rectangular layout \cite{clements2016optimal} shown in Fig. \ref{fig:setup}a. The device has a footprint of $80$ $\text{mm}^2$, and is fiber-pigtailed to single-mode fiber arrays both at the input and at the output facets. The device is characterized by an overall input-output transmission of $\sim50\%$, that takes into account also the fiber coupling of the input and output single-mode fiber array. Its operation can be changed by tuning $56$ phase parameters via application of electrical currents to the corresponding thermal shifters. A crucial aspect to exploit the device is the capability of correctly programming its operation. This requires a preliminary calibration procedure, that aims at obtaining a map to relate the currents in the resistors with the effective implemented transformation. The device calibration consists of finding all possible cross-talk coefficients so as to comprehensively characterize how each resistor affects all the waveguides on the chip. Once the calibration procedure is completed, its efficacy is tested by dialling Haar-random matrices, and then verifying the fidelity between the programmed transformations and the expected one. In our case, the average fidelity between the matrix moduli over dialled Haar-random matrices is found to be $F=0.995\pm0.002$.

\textbf{Overall $n$-fold coincidence rates.} Finally, the output states are measured via a set of high-efficiency superconducting nanowire single-photon detectors, with an average efficiency between the different channels of $\sim85\%$. Considering the detection efficiency, the generation rate and the overall transmission of the setup, we obtained $n$-fold detection rates integrated over all useful configurations of respectively $\sim5$ MHz for $n=1$, $\sim50$ kHz for $n=2$ and $\sim8$ Hz for $n=4$.

\subsection*{Neural network architectures}
The DAE was designed to recover the ideal data from the noisy ones, learning a non-linear mapping between corrupted input data and noiseless targets. The architecture comprises an encoder, a bottleneck, and a decoder. The encoder reduces the dimensionality of the input data, compressing it into a latent representation through two densely connected layers with 64 and 32 neurons, each employing a rectified linear unit (ReLU) activation function. The bottleneck layer at the core of the network consists of 16 neurons and is critical for capturing the most relevant features while filtering out noise. The decoder reconstructs the input data from the latent space, using two densely connected layers with 32 and 64 neurons, and a final output layer that matches the dimensionality of the input. A linear activation function in the output layer ensures that the reconstructed probabilities retain their original scale and structure. Once the probabilities are reconstructed, they are rescaled to ensure proper normalization. 

For phase estimation, a separate feed-forward NN was implemented to refine and linearize the estimation process. The NN maps the denoised conditional probabilities (eventually obtained at the output of the DAE) to the true phase values, addressing non-linear distortions and systematic biases in the raw estimation process. The network begins with an input layer of 16 neurons, corresponding to the $8$ conditional probabilities $p(\vec{m}|\phi)$, and $8$ probabilities $p(\vec{m}|\phi+\delta\phi)$, relative to an arbitrary (and fixed) shift $\delta\phi\simeq0.44$ rad. Including such additional information, enhances the network ability to resolve ambiguities, particularly in boundary regions where the periodicity of the function could otherwise cause confusion, as demonstrated in \cite{cimini2021calibration}. The NN architecture is then constituted of three hidden layers, each with 52 neurons, using sigmoid activation for the input layer and hyperbolic tangent (tanh) activation functions for the hidden layers. The final output layer consists of a single neuron with a linear activation function, providing a continuous phase estimate.

The NN was trained on a dataset generated from simulated measurement outcomes, corresponding to a wide range of phase values with added noise to reflect realistic experimental conditions. The network was optimized using the Adam optimizer and a mean squared error loss function, ensuring that the predicted phase values closely matched the true values. The training process involved 4000 epochs with 10 samples per batch.

Once trained, the NN was fed with the denoised conditional probabilities produced by the DAE, yielding refined phase estimates. This approach significantly reduced the impact of systematic deviations, enhancing robustness to experimental imperfections.

\section*{Acknowledgments}
This work was supported by ICSC – Centro Nazionale di Ricerca in High Performance Computing, Big Data and Quantum Computing, funded by European Union – NextGenerationEU, and by the ERC Advanced Grant QU-BOSS (QUantum advantage via nonlinear BOSon Sampling, grant agreement no. 884676). C.P. would like to acknowledge funding from the EU through the EIC Transition project FUTURE (grant agreement no. 101136471)



\end{document}


\title{Supplementary Information: Quantum Analog-to-Digital Converter for phase estimation}

\author{Eugenio Caruccio}
\affiliation{Dipartimento di Fisica, Sapienza Universit\`{a} di Roma, Piazzale Aldo Moro 5, I-00185 Roma, Italy}

\author{Simone Roncallo}
\affiliation{Dipartimento di Fisica, Universit\`{a} di Pavia, via A. Bassi 6, I-27100 Pavia, Italy}

\author{Valeria Cimini}
\affiliation{Dipartimento di Fisica, Sapienza Universit\`{a} di Roma, Piazzale Aldo Moro 5, I-00185 Roma, Italy}

\author{Riccardo Albiero}
\affiliation{Istituto di Fotonica e Nanotecnologie, Consiglio Nazionale delle Ricerche (IFN-CNR), Piazza Leonardo da Vinci, 32, I-20133 Milano, Italy}

\author{Ciro Pentangelo}
\affiliation{Ephos, Viale Decumano 34, I-20157 Milano, Italy}

\author{Francesco Ceccarelli}
\affiliation{Istituto di Fotonica e Nanotecnologie, Consiglio Nazionale delle Ricerche (IFN-CNR), Piazza Leonardo da Vinci, 32, I-20133 Milano, Italy}

\author{Giacomo Corrielli}
\affiliation{Istituto di Fotonica e Nanotecnologie, Consiglio Nazionale delle Ricerche (IFN-CNR), Piazza Leonardo da Vinci, 32, I-20133 Milano, Italy}

\author{Roberto Osellame}
\affiliation{Istituto di Fotonica e Nanotecnologie, Consiglio Nazionale delle Ricerche (IFN-CNR), Piazza Leonardo da Vinci, 32, I-20133 Milano, Italy}

\author{Nicol\`{o} Spagnolo}
\affiliation{Dipartimento di Fisica, Sapienza Universit\`{a} di Roma, Piazzale Aldo Moro 5, I-00185 Roma, Italy}

\author{Lorenzo Maccone}
\affiliation{Dipartimento di Fisica, Universit\`{a} di Pavia, via A. Bassi 6, I-27100 Pavia, Italy}

\author{Chiara Macchiavello}
\email{chiara.macchiavello@unipv.it}
\affiliation{Dipartimento di Fisica, Universit\`{a} di Pavia, via A. Bassi 6, I-27100 Pavia, Italy}

\author{Fabio Sciarrino}
\email{fabio.sciarrino@uniroma1.it}
\affiliation{Dipartimento di Fisica, Sapienza Universit\`{a} di Roma, Piazzale Aldo Moro 5, I-00185 Roma, Italy}

\maketitle

\section{Experimental noise in the photonic platform}
\label{SI:Input}

In this Section, we discuss the role and the effect on the estimation process of the main noise sources present in the experimental platform. More specifically, the different noise processes arise from imperfections in each experimental component. The main effects that need to be addressed are due to the photonic source and to the implementation of the different interferometers. For the first case, one needs to substantially consider the role of partial photon distinguishability and multi-photon components, which in both cases affect the preparation of the required GHZ resource states. Additionally, for the noise in the interferometer, one needs to take into account errors in the circuit programming, which affect all stages of the experiment (state preparation, phase encoding and measurement), and losses, the latter related to the multi-photon components arising from the source.

\subsection{Role of partial photon distinguishability}

The generation of the 2-photon and 4-photon states rely on genuine multiparticle interference between the input photons. For instance, the 4-photon resource state $\vert \tilde{\mathrm{GHZ}}_{4} \rangle = (\vert 0101 \rangle + \vert 1010 \rangle)/\sqrt{2}$ is generated in post-selection by quantum interference of the two different paths that photons can take in the interferometer. The presence of coherence between the two paths is thus related to the indistinguishability between the two different paths, and thus to the indistinguishability of the photons in all internal degrees of freedom (polarization, central wavelength and spectrum, ...). If only a single photon is distinguishable from the other three particles, the two paths can be in principle discriminated, thus leading to the generation of a density matrix with an incoherent mixture of the two terms $(\vert 0101 \rangle \langle 0101 \vert + \vert 1010 \rangle \langle 1010 \vert)/2$. Hence, the generation of the resource state is ultimately related to the genuine multiphoton indistinguishable component of the input photons obtained from the source. Analogous discussion is present for the generation of the 2-photon resource state $\vert \tilde{\mathrm{GHZ}}_{2} \rangle = (\vert 01 \rangle + \vert 10 \rangle)/\sqrt{2}$, while the 1-photon case is unaffected by this noise source. Considering a pure state model for the internal degrees of freedom, the indistinguishability of a $n$-particle multiphoton state is quantified by the Gram-matrix $S_{ij} = \langle \psi_{i} \vert \psi_{j} \rangle$, where $\vert \psi_{k} \rangle$ (with $k=1, \ldots, n$) describe the state of the internal degrees of freedom for photon $k$.

\begin{figure*}[ht!]
\includegraphics[width=0.99\textwidth]{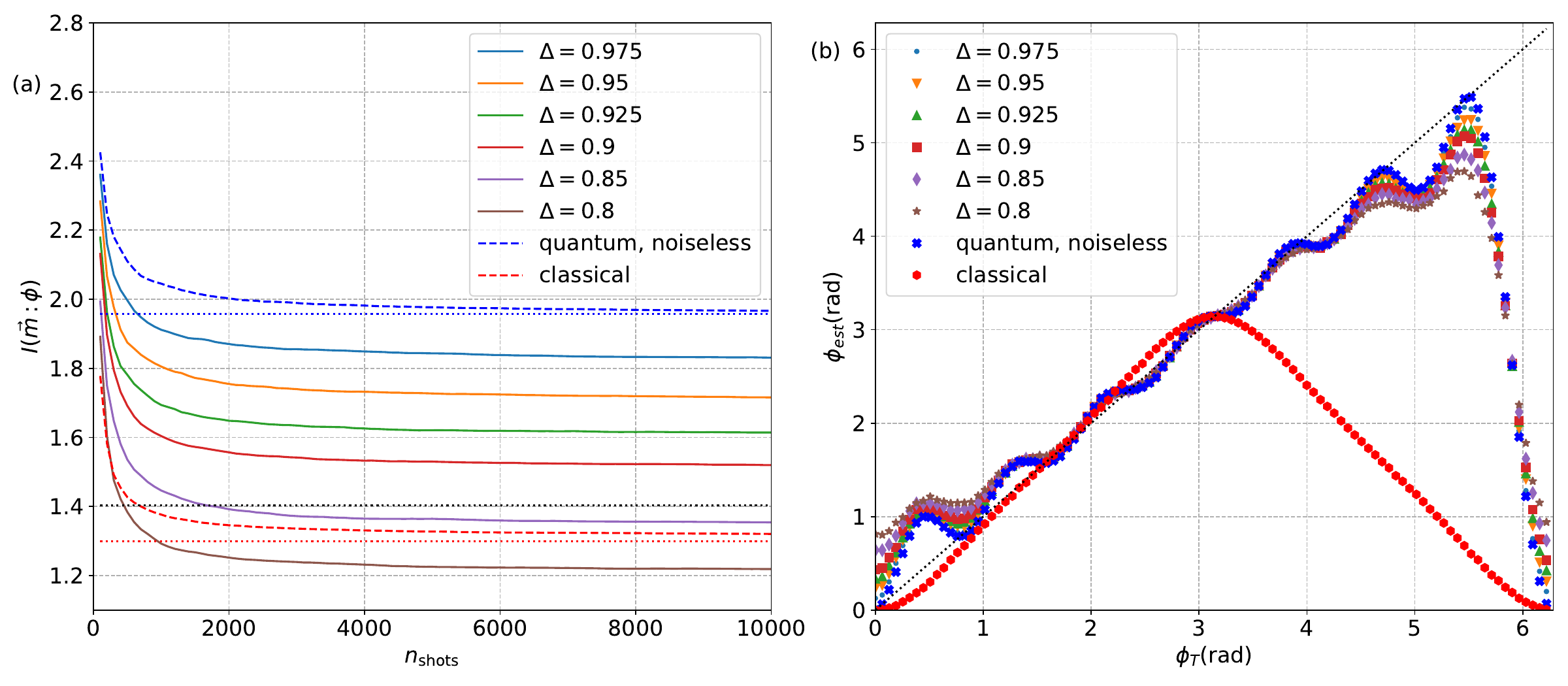}
\caption{\textbf{Simulation of the protocol with partial photon distinguishability.} Numerical simulation for $N=99$ equally spaced phase in the $[0, 2\pi)$ interval, $n_{\mathrm{shots}}$ repetitions for each phase, and different values of $\Delta$ (reported in the figure legend for each curve). (a) Mutual information $I(\vec{m}:\phi)$ as a function of $n_{\mathrm{shots}}$. For comparison, we report also numerical simulations for the quantum and classical strategies in absence of noise, and as horizontal lines the asympotic limit for the quantum strategy (dotted, blue), for the classical strategy (dotted, red) and the SQL (dotted, black). (b) Estimated phase from the protocol $\phi_{\mathrm{est}}$ as a function of the true value $\phi_{T}$, with $n_{\mathrm{shots}}=10^{4}$ repetitions per phase. For comparison, we report also numerical simulations for the quantum and classical strategies in absence of noise, and as a reference the black dotted line corresponding to $\phi_{\mathrm{est}} = \phi_{T}$.}
\label{fig:simulazioni_ind}
\end{figure*}

To provide a quantitative analysis of the role of this noise source, we have performed some numerical simulations as a function of the photon indistinguishability. In particular, we considered a real-valued Gram-matrix, a condition satisfied by our quantum-dot source \cite{rodari2024semi}, with a simplified model where the matrix has elements $S_{ij} = \sqrt{\Delta} + (1 - \sqrt{\Delta}) \delta_{ij}$, where $\delta_{ij}$ is the Kronecker delta. This scenario corresponds to the case where all particles are described by the same overlap $\vert \langle \psi_{i} \vert \psi_{j} \rangle \vert^{2} = \Delta, \forall (i,j)$. The parameter $\Delta$ is strictly related to the Hong-Ou-Mandel visibility \cite{olli21imperfect}. We have then performed some numerical simulations of the protocol for decreasing values of the parameter $\Delta$. The results are reported in Fig. \ref{fig:simulazioni_ind}a, where we show the curves for the mutual information as a function of the number of repetitions $n_{\mathrm{shots}}$ for $N=99$ different phases equally spaced in the full $[0, 2\pi)$ interval, as performed for the experimental implementation. We observe that, below a certain threshold $\Delta_{\mathrm{min}}$, this noise source provides a mutual information for the raw data which can fall below the value achievable with the classical protocol. Additional, in Fig. \ref{fig:simulazioni_ind}b we plot the curve of the estimated phases for the raw data as a function of $\Delta$, showing that the precision in the estimation is degraded for reduced indistinguishability between the particles. Even in ideal conditions ($\Delta = 1$), the estimated phase exhibits an oscillatory pattern. Such bias is intrinsic to the estimation strategy, regardless of noise or the number of experimental repetitions.

\subsection{Role of multiphoton components}

A second source of experimental noise arising from the source can be found in the presence of multiphoton components. More specifically, the quantum-dot source emits a train of single photons in different time-bins, separated by a fixed time distance depending on the repetition rate of the excitation pump. However, with a non-zero probability it is possible to observe the presence of a second photon in the same time-bin. As shown in Ref. \cite{olli21imperfect}, for the kind of source employed in the reported experiment this second photon is distinguishable from the other photons. In summary, on each time-bin the state generated by the source takes the form $\rho_{s} \sim p^{(0)} \vert \emptyset \rangle \langle \emptyset \vert + p^{(1)} \vert 1 \rangle \langle 1 \vert + p^{(2)} \vert (1,\tilde{1}) \rangle \langle (1,\tilde{1}) \vert$, where $p^{(0)}$, $p^{(1)}$ and $p^{(2)}$ are respectively the probabilities that 0, 1 or 2 photons are present in a single time-bin, and $(1,\tilde{1})$ stands for two photons in the mode, where the additional photons is in a distinguishable state of the internal degrees of freedom. The values of terms $p^{(1)}$ and $p^{(2)}$ \cite{pont2022quantifying} can be retrieved from a measurement of the second-order correlation function $g^{(2)}(0)$ via a Hanbury-Brown and Twiss interferometer, and by the value of the source brightness $B$.

Due to these multiphoton emission components, the $n$-photon state at the input of the interferometer includes additional terms. As an example, the principal component in the 4-photon case is given by an input state of the form $\vert 1_{1}, 1_{3}, 1_{5}, 1_{7}\rangle$, where $1_{i}$ stands for a photon in input port $i$ of the interferometer, with probability $[p^{(1)}]^4$. However, due to multiphoton emission, other possible combinations are present in the state generated from the source, such as terms of the form $\vert(1,\tilde{1})_{1}, 1_{3}, 1_{5}, 0_{7} \rangle$ (and permutations) with probability $[p^{(1)}]^2 p^{(0)} p^{(2)}$, or terms of the form $\vert(1,\tilde{1})_{1}, 1_{3}, 1_{5}, 1_{7} \rangle$ (and permutations) with probability $[p^{(1)}]^3 p^{(2)}$. These terms give rise to undesired contributions in the density matrix, which are mostly non-zero values in elements of the density matrix diagonal different from $\vert 0101 \rangle \langle 0101 \vert$ and $\vert 1010 \rangle \langle 1010 \vert$. Note that the weights of the different contributions in the full density matrix are also affected by the amount of losses $\eta$ in the apparatus, and by adoption of threshold detectors at the measurement stage. In the case of the employed apparatus, losses are almost balanced between the modes, and thus in the model they can be made to commute with the linear-optical components and placed entirely a the input of the device \cite{oszm18simulation}.

\begin{figure*}[ht!]
\includegraphics[width=0.99\textwidth]{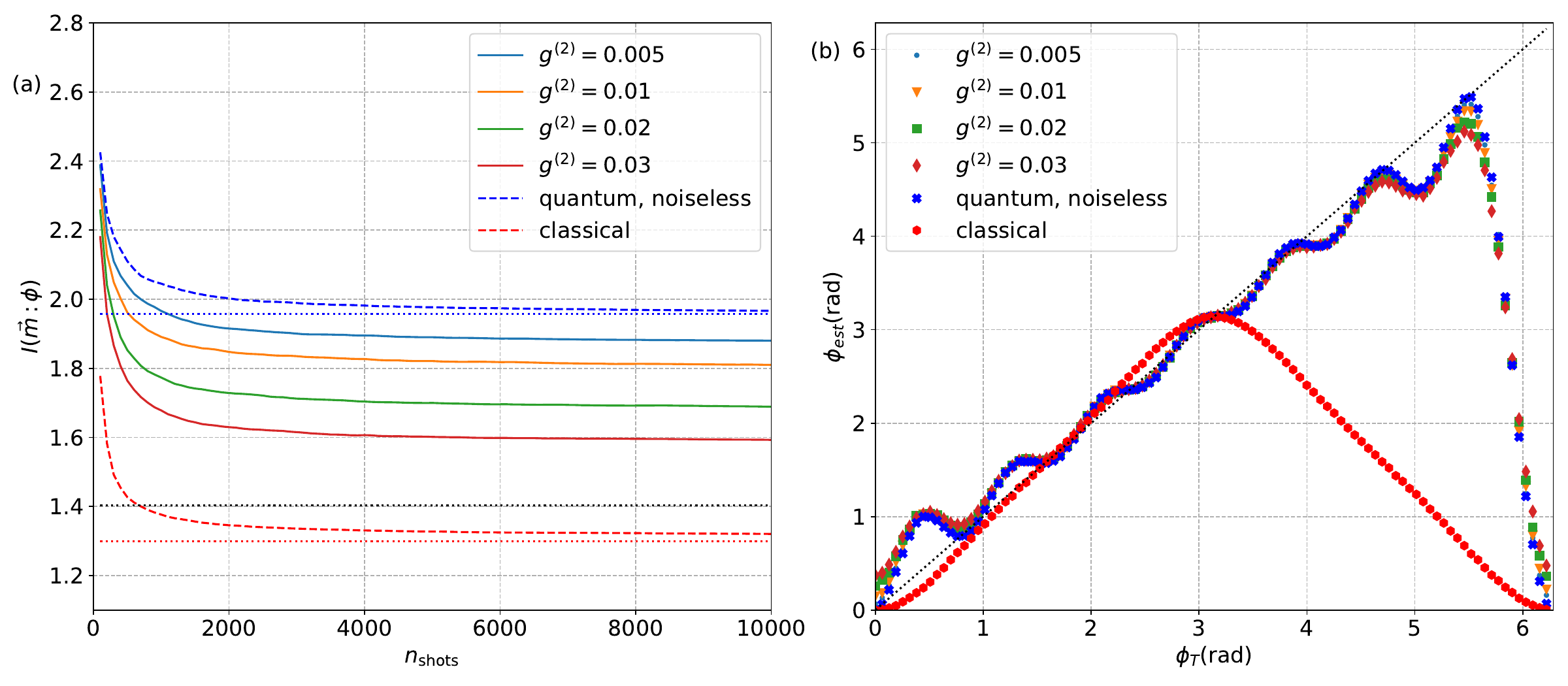}
\caption{\textbf{Simulation of the protocol with multiphoton components from the source.} Numerical simulation for $N=99$ equally spaced phase in the $[0, 2\pi)$ interval, $n_{\mathrm{shots}}$ repetitions for each phase, and different values of $g^{(2)}$ (reported in the figure legend for each curve). (a) Mutual information $I(\vec{m}:\phi)$ as a function of $n_{\mathrm{shots}}$. For comparison, we also report numerical simulations for the quantum and classical strategies in absence of noise, and as horizontal lines the asympotic limit for the quantum strategy (dotted, blue), for the classical strategy (dotted, red) and the SQL (dotted, black). (b) Estimated phase from the protocol $\phi_{\mathrm{est}}$ as a function of the true value $\phi_{T}$, with $n_{\mathrm{shots}}=10^{4}$ repetions per phase. For comparison, we report also numerical simulations for the quantum and classical strategies in absence of noise, and as a reference the black dotted line corresponding to $\phi_{\mathrm{est}} = \phi_{T}$.}
\label{fig:simulazioni_g2}
\end{figure*}

We have then carried out some numerical simulations to verify the impact of such multiphoton contributions in the estimation protocol. The results of this simulation are reported in Fig. \ref{fig:simulazioni_g2} for different values of the $g^{(2)}(0)$ parameter, and $N=99$ different phases in the $[0, 2\pi)$ interval as performed in the experiment. We observe, as expected, that this effect reduces the mutual information from the measured bit string, and that additional errors are introduced in the estimated phase $\phi_{est}$.

\subsection{Role of errors in circuit programming}

As a final relevant source of experimental noise, one needs to consider errors in programming the reconfigurable interferometer. More specifically, as shown in Fig. 2 of the main text, the different steps of the protocol are implemented by reconfiguring the operation of the integrated interferometer to perform each specific unitary transformation. This process requires a prior calibration of the device, namely to reconstruct a mapping between the applied currents in the resistors, and the corresponding values of the phase shifts on top of the waveguides. Hence, errors in the calibration process, and imperfect implementation of the static components, imply that slightly different unitaries than the desired ones are implemented by the interferometer. 

The reprogrammable phases (see Fig. 2a of the main text) can be divided in those ($\theta$) affecting the transmittivities $R(\theta)$ of the elementary cells, and those ($\varphi$) affecting the relative phases between the interferometer arms. Imperfections in the first set of parameters can affect the fidelity of the generated input resource, or introduce small errors in the Hadamard operations at the measurement stage. Imperfections in setting parameters $\varphi$ can introduce errors in other portion of the protocols, such as the insertion of the unknown phases $\phi$, or of the conditional $\sigma_z$ or $R^{-1}_{l}$ operations.

\section{Experimental raw data analysis}

In this Section we provide more details on the data analysis for the raw data before processing them via machine learning technique. More specifically, in Sec. \ref{subsec:processing} we describe the data processing approach to obtain the bit strings $\vec{m}$, according to the protocol, from the data collected by the full set of configurations. Then, in Sec. \ref{subsec:estimates} we describe and report the estimation of the phases from the collected strings, while in Sec. \ref{subsec:mutualinfo} we describe the evaluation of the mutual information from the measured $\vec{m}$.

\subsection{Data acquisition and processing}
\label{subsec:processing}

As mentioned in the main text, the data are gathered for all possible combinations of controlled operations $\sigma_z$ and $R_l^{-1}$. The employed post-processing, used in the platform to implement the different steps of the protocol, can be schematically described as a procedure of three main points: 1) in the $4-$ and $2$-photon experiments, a specific measured unitary outcome is discarded if the implemented controlled operation does not match the parity of the bit subset, based on the protocol rules. At the end, an array of bit strings, each associated with a run of the experiment, is produced for any controlled operation realization. 2) The successful results of the different measured unitary units are clustered in random order, retaining the information of the unitary from which they were derived, in three different arrays of $n$-bit strings, one for each experiment of $n$-photons, with $n=1,2,4$. Repetition of the protocol will involve the homologous components of these arrays. 3) All repetitions in which no agreement occurs between the controlled operation, implemented in one experiment, and the bit produced in another experiment, are discarded, according to the rules of the protocol. Post-proccessing yields a vector $\vec{m}$ for each of $99$ the phases $\phi_T$ and for each valid repetition according to the algorithm constraints. The vector $\vec{m}$ has dimension $7$ and contains the bits of, respectively, the $4$-, $2$- and $1$-photon experiment. 

\subsection{Estimated phases from the raw data}
\label{subsec:estimates}

After the data acquisition and processing stage described above, it is possible to provide a first estimate of the unknown phase from the raw data. In the quantum estimation protocol, for each phase and repetition providing a 7-bit string $\vec{m}$ the phase is estimated as $\phi_{\mathrm{est}} = 2 \pi (b_1/2 + b_2/4 + b_3/8)$, being $b_i$ the three relevant bits from $\vec{m}$. Then, for each value of the phase, repeating the experiment $n_{\mathrm{shots}}$ times provide $n_{\mathrm{shots}}$ independent estimates of $\phi_{\mathrm{est}}$. In the classical protocol, given the a 7-bit string $\vec{m}$, the phase is estimated as $\phi_{\mathrm{est}} = 2 \arccos \sqrt{p(0)}$, where $p(0)$ is the experimental estimate of the probability $p(\vert 0 \rangle \vert \phi)$ of obtaining the bit $0$. The experimental estimate is obtained as $p(0) = N_{0}/7$, where $N_{0}$ is the number of 0s in the string $\vec{m}$.

\begin{figure*}[ht!]
\includegraphics[width=0.99\textwidth]{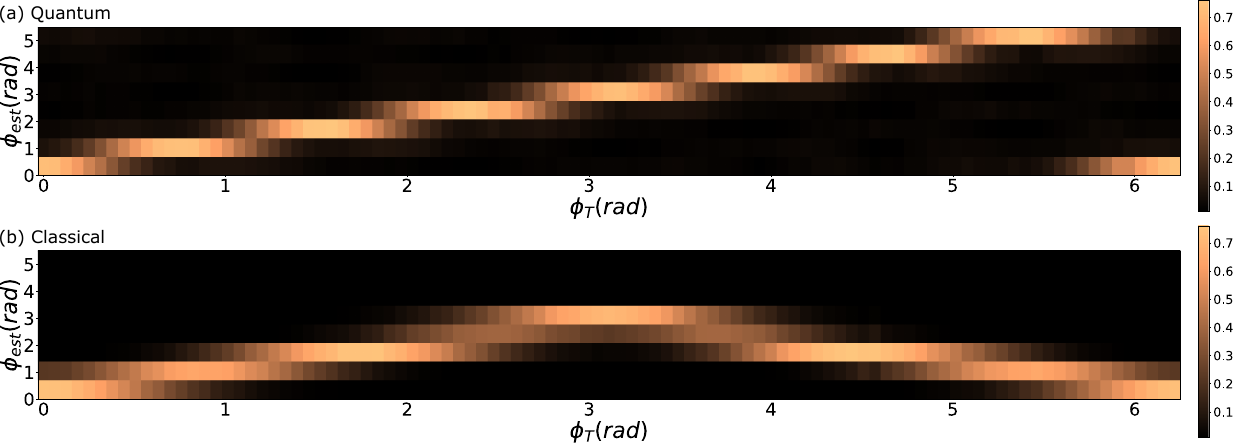}
\caption{\textbf{Histograms of the estimated phase $\phi_{\text{est}}$ for repeated shots of the experiment.} (a) Frequency of the eight possible values of $\phi_{\mathrm{est}} = 2 \pi (b_1/2 + b_2/4 + b_3/8)$ arising from the different combinations of the three bits $\vec{b} = (b_{1}, b_{2}, b_{3})$ for the quantum strategy. (b) Frequency of the estimated phase $\phi_{\mathrm{est}}$ with the classical strategy.}
\label{fig:isto fasi}
\end{figure*}

In Fig. \ref{fig:isto fasi}, we report the results of the raw experimental estimates of the phases obtained by applying respectively the quantum and classical protocol described in the main text. The protocols are repeated for $N=99$ different phases equispatially distributed in the $[0, 2\pi)$ interval, with $n^{\mathrm{max}}_{\mathrm{shots}}=5377$ repetitions for each phase. The histogram of the quantum protocol, in Fig. \ref{fig:isto fasi}a, shows the distribution of the measured phases, estimated from the single repetitions, for each phase $\phi_T$ (fixed column). The bins are centered on the 8 possible values (fixed row) of the estimated phase $\phi_{est}$. Each distribution is normalized with respect to $n^{\mathrm{max}}_{\mathrm{shots}}$ and the color is applied to display the histogram bar heights, taking the colormap on the side as a reference. The histogram in Fig. \ref{fig:isto fasi}b depicts the analogous result for the classical protocol considering the same binning arrangement as for the quantum case. The results presented in Fig. \ref{fig:isto fasi} complement the main text results, where the average values of the phases estimated from the raw data are shown. We conclude by observing that the oscillatory behaviour of the average estimated phases is intrinsic to the protocol, given that the marginal probabilities on the post-selected measurement outcomes contain a truncated set of Fourier harmonics in the phase $\phi$. This is also shown in the numerical simulations of Figs. \ref{fig:simulazioni_ind} and \ref{fig:simulazioni_g2} for the noiseless scenario.

\subsection{Evaluation of the mutual information from the raw data}
\label{subsec:mutualinfo}

As discussed in the main text, the number of useful bits is quantified in the protocol by the mutual information $I(\vec{m}:\phi)$ associated to a given bit string $\vec{m}$. This parameter is defined as:
\begin{equation}
I(\vec{m}:\phi) = \int_{0}^{2 \pi} d\phi \sum_{\vec{m}} p(\phi) p(\vec{m} \vert \phi) \log_{2}[p(\vec{m}\vert \phi)/p(\vec{m})].
\end{equation}
Here, $p(\phi)$ is the prior information on the phase, here assumed to be uniform $p(\phi) = 1/2\pi$, $p(\vec{m}\vert \phi)$ is the likelihood of obtaining a given string $\vec{m}$ for a given phase value, and $p(\vec{m})$ is the probability of obtaining $\vec{m}$. An estimate of $I(\vec{m}:\phi)$ from an experiment with $N$ phases and $n_{\mathrm{shots}}$ repetition can be obtained starting from the collected set of bit strings $\vec{m}$. In our case, the number $N$ of equally distributed phases is large enough so that the sampled phases can be representative of the uniform prior. The likelihood for each phase is obtained as $p(\vec{m}\vert \phi) = N_{\vec{m}}/n_{\mathrm{shots}}$, where $N_{\vec{m}}$ is the number of occurrence of a given string $\vec{m}$ among the $n_{\mathrm{shots}}$ experimental repetitions at fixed phase. Then, the probability $p(\vec{m}) = \int d\phi p(\phi) p(\vec{m}\vert \phi)$ is obtained by averaging the experimental estimates of $p(\vec{m}\vert \phi)$ over all sampled phases. Finally, errors in the experimental estimates of $I(\vec{m}:\phi)$ can be estimated by considering that the counting statistics follows a Poissonian distribution, and thus standard bootstrapping approaches can be applied.


%